\documentclass[12pt]{article}
\usepackage{amsmath, latexsym, amssymb, amscd, float,cite}
\usepackage{slashed, tikz}
\usetikzlibrary{arrows,shapes}
\usetikzlibrary{trees}
\usetikzlibrary{matrix,arrows} 				
\usetikzlibrary{positioning}				
\usetikzlibrary{calc,through}				
\usetikzlibrary{decorations.pathreplacing}  
\usepackage{pgffor}							

\usetikzlibrary{decorations.pathmorphing}	
\usetikzlibrary{decorations.markings}
\usetikzlibrary{shapes,arrows,positioning,automata,backgrounds,calc,er,patterns}

\restylefloat{table}
\newcommand{\C}{\mathbb{C}}
\newcommand{\R}{\mathbb{R}}

\newcommand{\bs}[1]{\mathbf{#1}}

\newtheorem{theorem}{Theorem}[section]

\newtheorem{proposition}[theorem]{Proposition}

\newtheorem{lemma}[theorem]{Lemma}

\newtheorem{postulate}[theorem]{Postulate}

\newtheorem{claim}[theorem]{Claim}
\newcommand{\QED}{\hspace{.2in}\square\newline}
\newcommand{\qED}{\hspace{.2in}\boxminus\newline}

\tikzset{
    vector/.style={decorate, decoration={snake}, draw},
	provector/.style={decorate, decoration={snake,amplitude=2pt}, draw},
	antivector/.style={decorate, decoration={snake,amplitude=-2pt}, draw},
    fermion/.style={draw=black, postaction={decorate},
        decoration={markings,mark=at position .55 with {\arrow[draw=black]{>}}}},
    fermionbar/.style={draw=black, postaction={decorate},
        decoration={markings,mark=at position .55 with {\arrow[draw=black]{<}}}},
    fermionnoarrow/.style={draw=black},
    gluon/.style={decorate, draw=black,
        decoration={coil,amplitude=4pt, segment length=5pt}},
    scalar/.style={dashed,draw=black, postaction={decorate},
        decoration={markings,mark=at position .55 with {\arrow[draw=black]{>}}}},
    scalarbar/.style={dashed,draw=black, postaction={decorate},
        decoration={markings,mark=at position .55 with {\arrow[draw=black]{<}}}},
    scalarnoarrow/.style={dashed,draw=black},
    electron/.style={draw=black, postaction={decorate},
        decoration={markings,mark=at position .55 with {\arrow[draw=black]{>}}}},
	bigvector/.style={decorate, decoration={snake,amplitude=4pt}, draw},
}
\begin{document}

\title{The Three Faces of $U(3)$}
\author{J. LaChapelle}

\maketitle

\begin{abstract}
$U(n)$ is a semi-direct product group that is characterized by non-trivial homomorphisms mapping $U(1)$ into the automorphism group of $SU(n)$. For $U(3)$, there are three non-trivial homomorphisms that induce three separate defining representations.  In a toy model of $U(3)$ Yang-Mills (endowed with a suitable inner product) coupled to massive fermions, this renders three distinct covariant derivatives acting on a single matter field. By employing a $\mathrm{mod}\,3$ permutation of the vector space carrying the defining representation induced by a ``large'' gauge transformation, the three covariant derivatives and one matter field can alternatively be expressed as a single covariant derivative acting on three distinct species of matter fields possessing the same $U(3)$ quantum numbers. One can interpret this as three species of matter fields in the defining representation.
\end{abstract}

\section{Introduction}
In this note we consider a toy model of $U(3)$ Yang-Mills coupled to massive fermionic matter fields. Off hand it seems $U(3)$ is an untenable symmetry group for constructing a gauge field theory. After all, a tenant of standard gauge theory says the most general symmetry group must be a direct product of semi-simple and $U(1)$ groups. (see e.g. \cite{W3})

From where comes the tenant? For a physically acceptable gauge field theory, one must start with a compact real group $G$ and impose a positive-definite, $Ad$-invariant, real bilinear form on the gauge symmetry Lie algebra $\mathfrak{g}$. And it is well-known that the Lie algebra of a compact real group decomposes into a direct sum of semisimple $\mathfrak{s}_i$ and $\mathfrak{u}(1)_j$ factors $\bigoplus_{i,j}\mathfrak{s}_i\oplus\mathfrak{u}(1)_j$ if and only if the Killing form on $\mathfrak{g}$ is non-degenerate and hence negative-definite (see e.g. \cite{FS}).

Because $U(3)$ is not semi-simple, its Killing form is degenerate. But a Killing inner product is not the only possibility. It happens that $U(3)$ is a connected, compact real group. Being compact, it is endowed with at least one bi-invariant metric\cite{GHV,CBDM}, and in fact one can formulate on $\mathfrak{u}(3)$ a two-parameter class of positive-definite, $Ad$-invariant, real bilinear forms. Hence, it is possible to construct a consistent gauge theory with $U(3)$ gauge symmetry without the Killing inner product.

Notably, unlike $SU(3)\times U(1)$ where the gauge field associated with $U(1)$ completely decouples from the rest, all of the $U(3)$ gauge fields will mutually interact as a true $U(3)$ symmetry dictates. In fact, we have $U(3)=SU(3)\rtimes U(1)$ as a semi-direct product, and an element $u_{(3)}\in U(3)$ can be factored as $u_{(3)}=s_{(3)}u_{(1)}$ with $u_{(1)}\in U(1)$ and $s_{(3)}\in SU(3)$. The semi-direct product $SU(3)\rtimes U(1)$ is characterized by a (not necessarily unique) homomorphism $\varphi:U(1)\rightarrow\mathrm{Aut}\,SU(3)$ where $\mathrm{Aut}\,SU(3)$ is the automorphism group of $SU(3)$.\cite{ROB, AP, KC} In particular, in the defining representation, said homomorphism induces a (not necessarily unique) representation $\varrho:U(1)\rightarrow L_B(\C^3)$ where $L_B(\C^3)$ denotes the set of linear bounded matrix operators on $\C^3$. Now, in the defining representation there are three \emph{non-trivial} ways to represent the $U(1)$ factor in $L_B(\C^3)$; with $e^{i\theta}$ in one of the diagonal entries, $1$ in the other two diagonal entries, and $0$ in all off-diagonal entries. Then an element of $U(3)$ represented in $L_B(\C^3)$ can be written $\rho_r(u_{(3)})=\rho_r(su_{(3)})\rho_r(u_{(1)})$ where $\rho_r:U(3)\rightarrow L_B(\C^3)$ is an extension of $\varrho_r$ and  $r\in\{1,2,3\}$.

There is no reason to favor one particular representation over another,  so when constructing a gauge field theory coupled to fermions in the defining representation the most general Lagrangian contains the standard Yang-Mills term $-\tfrac{1}{2}F\cdot F$ and fermion terms $\sum_r\overline{\Psi}\slashed{D}^{(r)}\Psi$ summed over the three representations $\rho_r$. Consider permuting some chosen basis of $\C^3$ with some unitary permutation matrix in $L_B(\C^3)$. There are two classes of such permutations: one class induces ``small'' gauge transformations and the other induces ``large'' gauge transformations. Of course, the small gauge transformations represent a redundant state description in the quantum version. In contrast, the large gauge transformations represent the non-abelian analog of charge conjugation, and they effect a genuine matter field re-characterization: They essentially permute $r$ up to non-trivial phases. Accordingly, the $U(3)$ symmetry allows the fermion contribution $\sum_r\overline{\Psi}\slashed{D}^{(r)}\Psi$ to be rewritten with the covariant derivative in  a single representation as $\sum_r\overline{\Psi}^{(r)}\slashed{D}\Psi^{(r)}$ where $\Psi^{(r)}$ are three \emph{different} species of fermion matter fields --- each species a $U(3)$ triplet characterized by three quantum numbers coming from the action of the Cartan subalgebra.

This is our main result: The most general $U(3)$ gauge invariant Lagrangian for fermions in a chosen defining representation includes precisely three species of matter fields relative to an imbedding $U(1)\hookrightarrow SU(3)\rtimes U(1)$. We make no claim here that $U(3)$ models QCD phenomenology.\footnote{In which case the $U(1)$ subgroup would have nothing to do with electromagnetism.} Our purpose is to point out the viability of semi-direct product groups in gauge field theories and to highlight the emanating effect of multiple defining representations: The three types of matter fields coming from $U(3)$ may or may not be a phenomenological red herring.

Of course, the occurrence of three generations in particle physics is still a mystery, and there have been attempts to explain the ``three'' using a variety of mechanisms. Most notable perhaps are preon models\cite{PS, T1, AH, DK}, and super string models \cite{GKMR, LS}. But there are also models based on non-anomalous discrete $R$-symmetry \cite{EIKY}, extra dimensions with anomaly cancellation \cite{DP}, and the anthropic principle \cite{IKY}.

\section{$U(3)$ Toy Model}
\subsection{The inner product}
$U(3)$ is neither simple nor semisimple, and its Killing form is only \emph{semi-definite}. So the first order of business is to construct a suitable inner product on $\mathfrak{u}(3)$.

\begin{proposition} The Killing form of $U(3)$ is given by $K(\mathbf{X},\mathbf{Y})
=6\mathrm{tr}(\mathbf{X}\cdot\mathbf{Y})
-2\mathrm{tr}(\mathbf{X})\mathrm{tr}(\mathbf{Y})$ and is negative semi-definite for all skew-Hermitian $\mathbf{X},\mathbf{Y}\in\mathfrak{u}(3)$.
\end{proposition}
\emph{Proof}: The Lie algebra brackets are $[\mathfrak{u}_{ab}\,,\,{\mathfrak{u}}_{cd}]
=\delta_{bc}\mathfrak{u}_{ad}-\delta_{ad}\mathfrak{u}_{cb}$ where $\mathfrak{u}_{ab}\in\mathfrak{u}(3)$ are a chosen skew-Hermitian basis with $a,b,c,d\in\{1,2,3\}$. From these brackets it follows that the adjoint map is given by ${ad}_{\mathbf{X}}(\mathfrak{u}_{ab})
=\sum_c x_{ca}\mathfrak{u}_{cb}-x_{bc}\mathfrak{u}_{ac}$ with $\mathbf{X}
=\sum_{a,b}x_{ab}\mathfrak{u}_{ab}$ and $x_{ab}\in\R$. Hence,
\begin{eqnarray}
{ad}_{\mathbf{X}}\circ{ad}_{\mathbf{Y}}(\mathfrak{u}_{ab})
&=&\sum_{c,d}(x_{ca}y_{dc}\mathfrak{u}_{db}-x_{bc}y_{da}\mathfrak{u}_{dc}
+x_{bc}y_{cd}\mathfrak{u}_{ad}-x_{ca}y_{bd}\mathfrak{u}_{cd})\notag\\
&=&\sum_{c}(x_{ca}y_{ac}-x_{bc}y_{aa}\delta_{cb}
+x_{bc}y_{cb}-x_{ca}y_{bb}\delta_{ac})\mathfrak{u}_{ab}
\end{eqnarray}
implies
\begin{eqnarray}
K(\mathbf{X},\mathbf{Y}):=\mathrm{tr}\left({ad}_{\mathbf{X}}\circ{ad}_{\mathbf{Y}}\right)
&=&\sum_{a,b,c}(x_{ca}y_{ac}-x_{bc}y_{aa}\delta_{cb}
+x_{bc}y_{cb}-x_{ca}y_{bb}\delta_{ac})\notag\\
&=&3\mathrm{tr}(\mathbf{X}\cdot \mathbf{Y})-\mathrm{tr}(\mathbf{X})\mathrm{tr}(\mathbf{Y})+3\mathrm{tr}(\mathbf{X}\cdot \mathbf{Y})-\mathrm{tr}(\mathbf{X})\mathrm{tr}(\mathbf{Y})\;.
\end{eqnarray}
The center of $\mathfrak{u}(3)$ is $\mathrm{span}_{i\R}\{\mathbf{1}\}$, and it is easy to see that $K(i\mathbf{1},\mathbf{X})=0$ for all $\mathbf{X}\in\mathfrak{u}(3)$. Negativity follows from the skew-hermiticity of $\mathbf{X},\mathbf{Y}$. $\QED$

This suggests to define a bilinear inner product on the Lie algebra $\mathfrak{u}(3)$ in the defining representation $\rho:U(3)\rightarrow L_B(\C^3)$ by
\begin{equation}\label{inner product}
  \langle\bs{\Lambda}_{\alpha},\bs{\Lambda}_{\beta}\rangle
  :=-\frac{1}{6}\left[6\,\mathrm{tr}(\bs{\Lambda}_{\alpha}\bs{\Lambda}^\dag_{\beta})
  -2\left(1-\frac{{g}_1^{2}}{{g}_2^{2}}\right)\mathrm{tr}(\bs{\Lambda}_{\alpha})
  \mathrm{tr}(\bs{\Lambda}^\dag_{\beta})\right]
\end{equation}
where the basis elements $\{\bs{\Lambda}_{\alpha}\}=\{\rho'(\mathfrak{u}_{ab})\}$ are $3\times3$ skew-Hermitian matrices with $\alpha\in\{1,\ldots,9\}$ and the parameters $g_1,g_2\in\R$ obey $0<g_2^2< g_1^2$. It is clearly positive-definite, $Ad$-invariant, and real. For a triangular decomposition of the basis $\{\bs{\Lambda}_{\alpha}\}$ denoted by $\{\mathbf{S}_a^\pm,\mathbf{H}_a\}$ with $a\in\{1,2,3\}$, the structure constants associated with the brackets $[\mathbf{S}_a^\pm,\mathbf{H}_a]$ differ from those associated with the Killing form. These structure constants, which are functions of $(g_1,g_2)$, characterize quantum numbers of non-neutral gauge bosons, and eigenvalues of $\{\mathbf{H}_a\}$ characterize quantum numbers of matter fields.

\subsection{Semidirect structure of $U(3)$}
Mathematically, it is fruitful to view $U(3)$ as an extension of a group $H\cong U(1)$ by a normal subgroup $N\cong SU(3)\vartriangleleft U(3)$. This is represented by the short exact sequence
\begin{equation}
1\longrightarrow N\stackrel{f}{\longrightarrow} U(3)\stackrel{\pi}{\longrightarrow} H\longrightarrow1\;.
\end{equation}
If there exists an \emph{injective homomorphism} $s:H\rightarrow U(3)$ such that $\pi\circ s=id_{H}$, then the extension is a semidirect product $N\rtimes H$. In this case, $U(3)$ can be regarded as a principle bundle with base $H$, structure group $N$, and global section(s) $s:H\rightarrow U(3)$. A choice of section corresponds to a choice of coset representative. Then $s(H)\cong U(1)\subset U(3)$ yields a unique decomposition $U(3)=SU(3)U(1)$ with $SU(3)\cap U(1)=\{id\}$, and $s$ induces a homomorphism $\tilde{\varphi}:s(H)\cong U(1)\rightarrow \mathrm{Aut}\,N$. These observations are demonstrated by the following theorem;

\begin{theorem}\emph{(\cite{ROB, AP, KC})}
Let $1\longrightarrow N\stackrel{f}{\longrightarrow} U(3)\stackrel{\pi}{\longrightarrow} H\longrightarrow1$ be a short exact sequence equipped with an injective homomorphism $s:H\rightarrow U(3)$ such that $\pi\circ s=id_{H}$. Then there exists a homomorphism $\varphi:H\rightarrow \mathrm{Aut}\,N$ and an isomorphism $\theta:U(3)\rightarrow N\rtimes_\varphi H$.
\end{theorem}

\emph{Proof}:
For $h\in H$ and $n\in N$,
\begin{equation}
\pi(s(h)f(n)s(h^{-1}))=\pi\circ s(h)\,\pi\circ f(n)\,\pi\circ s(h^{-1})
=h\,id_H\,h^{-1}\;.
\end{equation}
Since $f$ is injective and $\mathrm{im}\,f=\mathrm{ker}\,\pi$, then $s(h)f(n)s(h^{-1})=f(n')$ for some unique $n'(n,h)\in N$ that depends on $(n,h)$. It is convenient to write $\varphi_h(\cdot)\equiv n'(\cdot,h)$ so that $\varphi_h:N\rightarrow N$. Note that $\varphi_h(id_N)=id_N$ for all $h\in H$ since $s$ is a homomorphism.

\begin{lemma}
The function $\varphi_h\in\mathrm{Aut}\,N$.
\end{lemma}

\emph{proof}:
First, $f(\varphi_{id_H}(n))=f(n)$ implies $\varphi_{id_H}(n)=n$ for all $n\in N$. Next, for $n_1,n_2\in N$,
\begin{equation}
s(h)f(n_1)f(n_2)s(h^{-1})
=s(h)f(n_1)s(h)^{-1}s(h)f(n_2)s(h^{-1})
=f(\varphi_{h}(n_1)\varphi_{h}(n_1))
\end{equation}
where we used $s$ is a homomorphism. On the other hand, from the definition of $\varphi_h$, we have $s(h)f(n_1n_2)s(h)^{-1}=f(\varphi_{h}(n_1n_2))$. Injective $f$ then implies $\varphi_{h}(n_1n_2)=\varphi_{h}(n_1)\varphi_{h}(n_2)$.
$\qED$

\begin{samepage}
Let $\varphi:H\rightarrow\mathrm{Aut}\,N$ by $h\mapsto\varphi_h$.
\begin{lemma}
$\varphi:H\rightarrow\mathrm{Aut}\,N$ is a homomorphism.
\end{lemma}
\end{samepage}
\emph{proof}: For $h_1,h_2\in H$,
\begin{equation}
 s(h_1)s(h_2)f(n)s(h_2)^{-1}s(h_1)^{-1}
 =s(h_1)f(\varphi_{h_2}(n))s(h_1)^{-1}
 =f(\varphi_{h_1}(\varphi_{h_2}(n)))\;.
\end{equation}
On the other hand, $s(h_1)s(h_2)f(n)s(h_2)^{-1}s(h_1)^{-1}=s(h_1h_2)f(n)s(h_1h_2)^{-1}=f(\varphi_{h_1h_2}(n))$ since $s$ is a homomorphism. Again injective $f$ implies $\varphi_{h_1h_2}=\varphi_{h_1}\circ\varphi_{h_2}$. $\qED$

It follows that $\varphi$ defines a group operation on $N\rtimes_\varphi H$ by $(n_1,h_1)(n_2,h_2)=(n_1\varphi_{h_1}(n_2),h_1h_2)$ if the inverse is defined by $(n,h)^{-1}:=(\varphi_{h^{-1}}(n^{-1}),h^{-1})$ for all $(n,h)\in N\rtimes_\varphi H$.

Finally, let $\theta^{-1}:N\rtimes_\varphi H\rightarrow U(3)$ by $(n,h)\mapsto f(n)s(h)$. Then
\begin{eqnarray}
\theta^{-1}((n_1,h_1)(n_2,h_2))
&=&\theta^{-1}(n_1\varphi_{h_1}(n_2),h_1h_2)\notag\\
&=&f(n_1)(s(h_1)f(n_2)s(h_1)^{-1})s(h_1)s(h_2)\notag\\
&=&f(n_1)s(h_1)f(n_2)s(h_2)\notag\\
&=&\theta^{-1}(n_1,h_1)\theta^{-1}(n_2,h_2)\;.
\end{eqnarray}
Since the decomposition $U(3)=NH$ is unique (which we won't bother to prove), the homomorphism $\theta^{-1}$ is bijective. One can go on to show that the semidirect product reduces to a direct product if and only if $H\vartriangleleft U(3)$; in which case $N$ and $H$ commute and $\varphi$ is trivial.$\QED$

Observe the homomorphism $\tilde{\varphi}:s(H)\cong H\in
N\rtimes_\varphi H \rightarrow\mathrm{Aut}\,N$ induced by $s$ is given by
\begin{eqnarray}
\tilde{\varphi}_{s(h)}(n,id_H)=s(h)(n,id_H)s(h^{-1})
&=&[(id_N,h)(n,id_H)](id_N,h^{-1})\notag\\
&=&[(\varphi_h(n),h)](id_N,h^{-1})\notag\\
&=&(\varphi_h(n),id_H)\;.
\end{eqnarray}
In this sense, $\tilde{\varphi}$ induced by the section $s$ coincides with $\varphi$. It is important to note that there may be multiple homomorphisms $\varphi$ and hence multiple sections $s$ that render a semidirect product. Physically, a non-trivial $\varphi$ corresponds to a direct interaction between the gauge fields of the respective subgroups.

In particular, for the matrix group $U(3)$ as a semidirect product, there exist three such non-trivial sections;
\begin{equation}
  s:H\rightarrow\left\{\begin{array}{c}
                  \mathrm{diag}(e^{i\omega},1,1) \\
                   \mathrm{diag}(1,e^{i\omega},1) \\
                   \mathrm{diag}(1,1,e^{i\omega})
                \end{array}\right.
\end{equation}
where $\omega\in\R$. Each section gives rise to a different conjugation of $SU(3)$ by $s(h)$, and each of these induces a different representation $\varrho_r:H\rightarrow L_B(\C^3)$ where $r\in\{1,2,3\}$. These can then be extended to three defining representations $\rho_r:U(3)\rightarrow L_B(\C^3)$.

\subsection{Lagrangian matter field term}
Given the existence of a suitable inner product and three representations, constructing the model is rather elementary. The decisive step is to insist that all allowed defining representations be included in the Lagrangian;
\begin{postulate}\label{matter representations}
The matter field portion of the Lagrangian of a gauge field theory must include all allowed defining representations.
\end{postulate}
Accordingly, in our toy model of Yang-Mills coupled to a massive matter field in the defining representations, the gauge field term uses the chosen inner product  $\tfrac{1}{2}\langle F,F\rangle$ with $F\in\mathfrak{u}(3)$ and the matter field term will be $\sum_ri\overline{\Psi}\slashed{D}_B^{(r)}\Psi$ where we have (unconventionally) included the bare mass parameter in the covariant derivative $\slashed{D}_B^{(r)}$. In momentum space, the matrix representation of the covariant derivative is $[\slashed{D}_B^{(r)}]=(\slashed{p}+m^{(r)}_B)[\mathbf{1}]
+\slashed{A}_B^\alpha[i\mathbf{\Lambda}^{(r)}_\alpha]$ with gauge fields $A_\mu^\alpha$, and $\{\mathbf{\Lambda}^{(r)}_\alpha\}$ a basis of $\mathfrak{u}(3)$ in the $r$-defining representation.

In the quantum version of this model, each $\slashed{D}^{(r)}$ will give rise to different vertex factors in the Feynman rules and hence \emph{ostensibly different} renormalizations of the gauge fields, matter fields, and $r$-dependent mass parameters. The renormalized matter field term is then $\sum_ri\overline{\Psi}\slashed{D}_R^{(r)}\Psi$ where $[\slashed{D}_R^{(r)}]
=(i\slashed{\partial}+m^{(r)}_R)[\mathbf{1}]+\slashed{A}_R^\alpha[i\mathbf{\Lambda}^{(r)}_\alpha]$. In effect, through renormalization, the quantum theory distinguishes the classically isomorphic vector spaces carrying the defining representations even when $m^{(r)}_B=\tfrac{1}{3}m_B$ where $m_B:=\sum_rm^{(r)}_B$. Notably, if there exists any bare mass degeneracy among the defining representations, the quantum version will remove the degeneracy (assuming different renormalizations for different $r$).

We can make use of the $U(3)$ symmetry to re-characterize the matter field Lagrangian. There exists a class of elements in $U(3)$ of the form
\begin{equation}
P(x):=\left(
     \begin{array}{ccc}
       0 & 0 & e^{i\theta_1(x)} \\
       e^{i\theta_2(x)} & 0 & 0 \\
       0 & e^{i\theta_3(x)} & 0 \\
     \end{array}
   \right)
\end{equation}
with $\theta_1(x),\theta_2(x),\theta_3(x)\in\R$. The adjoint action of $P(x)$ on the Lie algebra $\mathfrak{u}(3)$ leaves the normal subalgebra $\mathfrak{su}(3)$ invariant, but it cyclically permutes the generators of the $s(H)$ matrices
\begin{equation}
\mathrm{diag}({i\omega},0,0)\stackrel{Ad_P}{\rightarrow}\mathrm{diag}(0,{i\omega},0)
\stackrel{Ad_P}{\rightarrow}\mathrm{diag}(0,0,{i\omega})
\stackrel{Ad_P}{\rightarrow}\mathrm{diag}({i\omega},0,0)\;.
\end{equation}
Similarly, $P^{-1}(x)=P^\dag(x)$ permutes in the reverse direction. Crucially, $P^3=e^{i(\theta_1+\theta_2+\theta_3)}\mathrm{diag}(1,1,1)$.
We claim that $\theta_1(x)+\theta_2(x)+\theta_3(x)=\pm (2n)\pi$ with $n\in\mathbb{N}$
induces small gauge transformations while $\theta_1(x)+\theta_2(x)+\theta_3(x)=\pm (2n+1)\pi$ induces large gauge transformations: The latter cannot be reached by a gauge transformation homotopic to the identity  because $\det{P}=-1$.\footnote{To see this use the identity in three dimensions $\det A=1/6\left((\mathrm{tr}\,A)^3-3\mathrm{tr}\,A\,\mathrm{tr}\,(A^2)+2\mathrm{tr}\,(A^3)\right)$ and put $A\rightarrow U(x)$ with $U(x)=\bs{1}+i\sigma^\alpha(x)\bs{\Lambda}_\alpha+O(\sigma^2)$ an infinitesimal gauge transformation. To first order in $\sigma$, find that $\det U(x)>0$.}  It then follows from $\mathrm{tr}\log P= i\pi(2k+1)$ that $\log P$  in this case involves a combination of Cartan generators (not present in the small permutation case) that contributes a \emph{multivalued} $\mathrm{mod}\,3$ phase to matter field configurations, and it transforms between three physically distinct classes of gauge field configurations that survive gauge fixing in the quantized theory.

Given $P$ we have $\slashed{D}^{(2)}=P\slashed{D}^{(1)}P^{-1}$ and $\slashed{D}^{(3)}=P^{2}\slashed{D}^{(1)}P^{-2}$. Define the fields $\Psi^{(r)}:=P^{r-1}\Psi$. Clearly $P$ cyclically permutes the components of $\Psi$ up to phases. Consequently, when $P^3=-\mathbf{1}$ we can write $\sum_r\overline{\Psi}\slashed{D}_B^{(r)}\Psi=\sum_r\overline{\Psi}^{(r)}\slashed{D}_B\Psi^{(r)}$. In the quantum version, if  the $\theta_i(x)$ are not all equal, the $SU(3)$-identical $\Psi^{(r)}$ are physically distinct fields with inequivalent masses (again, assuming different renormalizations for different $r$). Hence we claim
\begin{claim}
Given postulate \emph{\ref{matter representations}}, matter fields with $U(3)$ gauge symmetry necessarily come in three species due to the existence of large gauge transformations that realize $\mathrm{mod}\,3$ permutations of the basis in a defining representation.
\end{claim}
This perspective can be turned around: One can view fermions in the defining representation as a single field, and different fermion species are just a manifestation of the three faces of $U(3)$.

\section{Summary and Outlook}
Our analysis started with the observation that $U(1)$ gauge symmetry can be incorporated into gauge field theories via semi-direct products and not simply as direct products. In particular, for $U(3)$ the construction of the semi-direct product is not unique; it comes in three versions. We argued these three versions can be interpreted as three species of matter fields. The interpretation relies on including all three versions of the semi-direct product in the Lagrangian, the large-gauge-transformation status of certain permutation operators, and the identification $\Psi^{(r)}:=P^{r-1}\Psi$.\footnote{We did not consider $U(2)$ as a replacement for $SU(2)\times U(1)$, but off hand the same mechanism would appear to apply and it should be studied in the context of spontaneous symmetry breaking.}

From here, it is natural to wonder if there could be realistic strong-force phenomenology coming from gauged $U(3)$. Long ago Fischbach, et. al.\cite{F1} proposed the symmetry group $SU(3)_{C}\times U(1)_B$ with $SU(3)_C$  being the color symmetry of QCD and  $U(1)_B$ coupling to baryon number, but it was effectively falsified by experiment\cite{F2}. However, the gauge-field interactions for $U(3)$ differ considerably from the $SU(3)\times U(1)$ case. All of the gauge fields associated with the Cartan subalgebra of $U(3)$ take part in both gauge and matter field interactions.\footnote{To remind, the $U(1)$ subgroup is \emph{not} the electromagnetic gauge symmetry.} So if there is somehow any vestige of a long-range charge carrier coming from $U(3)$, it will couple to both gauge and matter field mass-energy and therefore have a chance of being consistent with gravity --- which was the downfall of $SU(3)_{C}\times U(1)_B$. Less clear and more imperative is whether $U(3)$ can somehow agree with QCD. 

There are reasons to suspect there might be some kind of charge-carrying abelian gauge field beyond the Standard Model. Along these lines, many models incorporate a ``dark photon'' that interacts with a hidden matter field sector but may or may not interact with the Standard Model sector. The dark photon literature is quite extensive: For a review see \cite{FGL} and references therein. The idea of appending a hypercolor symmetry group $SU(3)_H\times U(1)_H$ to the minimal  supersymmetric $SU(5)_{GUT}$ was studied in \cite{HIY, IY, IWY, WY}. The extra factor group\footnote{The authors of \cite{IWY} and \cite{WY} often write $U(3)_H$ but they put $U(3)_H\equiv SU(3)_H\times U(1)_H$ in \cite{IWY} and remark that $U(3)_H\simeq SU(3)_H\times U(1)_H$ in \cite{WY}. In both papers they use $SU(3)_H\times U(1)_H$ in calculations.} resolves some shortcomings of the model, and it can be viewed as a $\mathrm{D}3-\mathrm{D}7$ brane system in type IIB supergravity. The semi-direct product group $(SU(3)_C\times SU(2)_L)\rtimes U(1)_Y$ and anomaly cancellation were used by \cite{HMM} to put constraints on matter field hypercharge. As evidenced by the literature, it is possible to construct interesting models with extra $U(1)$ factors. We hope this note together with these studies will spur further investigation.

\end{document}